\documentclass[preprint,
normalem,
eqsecnum,
amsmath,amssymb,
amsmath,amssymb, aps,
]{revtex4}

\usepackage[dvipsnames]{xcolor}
\usepackage{epsfig}
\usepackage{amsfonts}
\usepackage{amsmath}
\usepackage{wasysym}
\usepackage{amssymb}
\usepackage{bm}
\usepackage{float}
\usepackage{enumitem}

\usepackage{graphicx}
\usepackage{ulem}

\usepackage[utf8]{inputenc}
\usepackage{bbm} 				% identity operator
\usepackage{epstopdf}				% self explanatory
\usepackage{verbatim}    			% long comments
\usepackage{sidecap}                % captions on the side
\usepackage{indentfirst}

\usepackage{nameref}

\def\mbf#1{\mathbf{#1}}
\def\bs#1{\boldsymbol{#1}}

\newcommand{\be}{\begin{equation}}
\newcommand{\ee}{\end{equation}}
\newcommand{\bea}{\begin{eqnarray}}
\newcommand{\eea}{\end{eqnarray}}

\usepackage[dvipsnames]{xcolor}
\usepackage{graphicx}% Include figure files
\usepackage{dcolumn}% Align table columns on decimal point
\usepackage{bm}% bold math

\usepackage[%
  colorlinks=true,
  urlcolor=blue,
  linkcolor=blue,
  citecolor=blue
]{hyperref}

\usepackage{soul}

\begin{document}

\bibliographystyle{apsrev4-2}

%\preprint{APS/123-QED}

%\title{On the memory effects of turbulent pipe flow through\\ the Lattice Boltzmann method}
\title{Persistence of large scale coherent structures in a turbulent pipe flow through an improved lattice Boltzmann approach}

\author{B. Magacho$^{1}$\footnote{bruno.magacho@impa.br}, L. Moriconi$^2$, and
J. B. R. Loureiro$^3$}
\affiliation{$^1$Instituto de Matemática Pura e Aplicada - IMPA, CEP: 22460-320, Rio de Janeiro, RJ, Brazil,}
\affiliation{$^2$Instituto de F\'\i sica, Universidade Federal do Rio de Janeiro,
C.P. 68528, CEP: 21941-972, Rio de Janeiro, RJ, Brazil,}
\affiliation{$^3$Programa de Engenharia Mecânica, Coordenação dos Programas de Pós-Graduação em em Engenharia, 
Universidade Federal do Rio de Janeiro, \\ {C.P. 68503, CEP: 21945-970, Rio de Janeiro, RJ, Brazil}}

\begin{abstract}
We simulated a turbulent pipe flow within the Lattice Boltzmann Method using a multiple-relaxation-time collision operator with Maxwell-Boltzmann equilibrium distribution expanded, for the sake of a more accurate description, up to the sixth order in Hermite polynomials. The moderately turbulent flow ($Re_{\tau} \approx 181.3$) is able to reproduce up to the fourth statistical moment with great accuracy, compared with other numerical schemes and with experimental data. A coherent structure identification was performed based on the most energetic streamwise turbulent mode, which revealed a surprising memory effect related to the large scale forcing scheme used to trigger the turbulent state in the pipe. We observe that the existence of large scale motions which are out of the pipe's stationary regime do not affect the detailed single-point statistical features of the flow. Furthermore, the transitions between the coherent structures of different topological modes were analyzed as a stochastic process. We find that for finely resolved data the transitions are effectively Markovian, but for larger decimation time lags, due to topological mode degeneracy, non-Markovian behavior emerges, in agreement with previous experimental studies.

\end{abstract}

%\keywords{Suggested keywords}%Use showkeys class option if keyword
                              %display desired
\maketitle

%\tableofcontents

\section{\label{sec:level1}INTRODUCTION}

The dynamics of turbulent flows is one of the most intriguing phenomena in classical physics. Concerning a broad set of applications turbulence is described by the Navier-Stokes (NS) equations of motion, for which, the existence of a unique and smooth solution is one of the millennium prize problems in mathematics \cite{Fefferman2006}. Till nowadays, to a great extent, the literature predictions of turbulent flows are associated with statistical features, such as the investigation of the turbulent kinetic energy, and statistical moments of velocity fluctuations, not to mention many others \cite{Frisch,moriconi2009}.

Coherent structures (CS) in turbulent flows seems to be part of the foundations of the physical background behind the output of statistical features. In wall-bounded turbulence, the CS can be related to quasi-streamwise vortices, hairpin vortices, large scales of motion (LSM), and very large scales of motion (VLSM) \cite{adrian2007,dennis2014,Balakumar2007}. Their existence was already identified experimentally with particle image velocimetry (PIV) \cite{adrian2007,dennis2014,Prasad1993,JackelPOF2023} and numerically by direct numerical simulations (DNS) with several structure detection mechanisms, such as the proper orthogonal decomposition (POD), dimensionality reduction, a local criterion depending on the vorticity field, and from the dynamical systems point of view \cite{Hellstrom2017,Schneider2007,Marensi2021,Elsas2017}.

In turbulent pipe flows, these CS could be assigned to a discrete azimuthal wave number, which is related to the most energetic streamwise mode \cite{Schneider2007, dennis2014}. From this perspective, a turbulent state --instantaneous snapshot-- could be thought as a superposition of different azimuthaly symmetric contributions from the turbulent streamwise kinetic energy. The aforementioned CS detection by dimensionality reduction opens the path to the study of the CS dynamics as a stochastic process \cite{JackelPRF2023}, which could bring important information about the mean lifetime of turbulent structures and the explicit dependence of individual mode contributions to desired observables.

In order to numerically simulate the NS equations, one can make use of several numerical schemes, such as finite differences, finite element, finite volume, and pseudospectral, among others. A competitive approach is the lattice Boltzmann method (LBM), due to its locality, it turns out to be a scalable and highly parallel alternative to simulate the NS equations \cite{Korner2006,Touil2016,Schornbaum2016,Xu_2018,LATT2021334,chen_doolen1998,cercignanibook}. Based on a discretized version of the Boltzmann equation, the LBM is a mesoscopic approach that was successfully applied to a variety of physical systems, such as magnetohydrodynamic (MHD) flows \cite{Magacho2023,Tavares2023,DELLAR2002}, thermal convection-diffusion \cite{Yoshida2010,Li2013}, multiphase flow \cite{Tavares2021}, and more.

The potential scalability of the LBM and its current development --with advanced collision operators and an accurate expansion of the Maxwell-Boltzmann distribution, to be discussed later-- increases the numerical stability of LBM simulations. The method has been applied to simulate wall-bounded turbulent flows (channels and pipe) and has shown to be an excellent alternative and easy-to-implement algorithm that gives with reasonable accuracy the same outputs from the aforementioned schemes, such as statistical moments of velocity fluctuations, vorticity, and Reynolds stress, among others \cite{Kang2013,Peng2018,Nathen2018}.

\section{\label{Methods}Methods}
\subsection{\label{LBM_METHOD_subsec}The Lattice-Boltzmann Method}

The LBM can be described straightforwardly as the mesoscopic time evolution of discretized particle distributions $f_i$ by a two-step algorithm, which, in lattice units and in the presence of an external force, is given by
\bea
    && f^*_i(\bs{x},t) = f_i(\bs{x},t) + \bs{L}[f_i(\bs{x},t)] + \left ( \mbf{I}-\frac{\bs{\Lambda}}{2} \right) \mathcal{F}_i  \ , \ \label{collision} \\
    && f_i(\bs{x} + \bs{c}_i,t+1) = f^*_i(\bs{x},t) \ , \label{streaming}
\eea

\noindent where $\bs{c}_i$ is the discretized lattice velocity, $\bs{L}$ is the collision operator, $\bs{\Lambda}$ is the collision matrix, and $\mathcal{F}_i$ takes into account the external forcing. Equations (\ref{collision}) and (\ref{streaming}) are known as the collision and stream steps, respectively.

Macroscopic quantities as the fluid density $\rho$ and the velocity field $\mathbf{u}$ are obtained as the zero and first order moments of distribution $f_i$, which are given by
\begin{eqnarray}
&&\rho = \sum_i f_i \ , \ \label{macroscopic_rho} \\
&&\mathbf{u} = \frac{1}{\rho}\sum_i \mathbf{c}_i f_i + \frac{\mathbf{F}}{2\rho}  \ , \
    \label{macroscopic_u}
\end{eqnarray}
\noindent where $\mathbf{F}$ is an arbitrary external body force.

The BGK approximation \cite{BGK1954}, also known as the single-relaxation-time (SRT) model, relax all the discretized distributions to a local equilibrium --given by Maxwell Boltzmann distribution-- by the same relaxation time $\tau$, it reads as
\be
    \boldsymbol{L}[f_i(\boldsymbol{x},t)] \equiv -\frac{1}{\tau}(f_i - f_i^{eq}) \ . \  \label{srt}
\ee

Although the SRT collision model seems straightforward, it lacks stability for small values of relaxation time, which is associated with higher values of the Reynolds number. This issue can be solved by the adoption of a more refined collision model, as the multiple-relaxation-time (MRT), which changes the space of the distributions $f_i$ to the space of moments $m_i$ and relax, with a relaxation time related to the viscosity, only the moments which are associated with the Reynolds stress \cite{dhumieresarticle1992,coveney_succi_dhumieres_ginzburg2002,hosseini2019,coreixas_chopard_latt2019,coreixas_wissocq_chopard_latt2020,krugerbook,lallemand_Luo2000}. A detailed explanation of the MRT model is given in Appendix \ref{Appendix}.

The no-slip boundary condition ($\mbf{u}\vert_{wall} = 0$) was implemented by the second-order accurate Boudizi scheme \cite{bouzidi2001}, which is a linear interpolated version of the bounce-back scheme \cite{Ziegler1993, Ginzbourg1994} and is given by
\bea
    &&f_{\overline{i}}(\boldsymbol{x}_f,t+1) = \frac{1}{2\Delta}\tilde{f}_{i}(\boldsymbol{x}_f,t) + \frac{2\Delta - 1}{2\Delta}\tilde{f}_{\overline{i}}(\boldsymbol{x}_f,t)  \mbox{   for  } \Delta \geq \frac{1}{2} \ , \  \label{fbc1}\\
    &&f_{\overline{i}}(\boldsymbol{x}_f,t+1) = 2\Delta \tilde{f}_{i}(\boldsymbol{x}_f,t) + (1 - 2\Delta) \tilde{f}_{i}(\boldsymbol{x}_{ff},t)  \mbox{   for  } \Delta < \frac{1}{2} \ , \ \label{fbc2}
\eea
where $\Delta$ is the fractional distance of the wall from the near-wall lattice point, $\boldsymbol{x}_f$ represent the fluid node, $\boldsymbol{x}_{ff} \equiv \boldsymbol{x}_{f} - \mathbf{c}_{i}$, $\overline{i}$ denotes the direction opposite to $\boldsymbol{c_i}$, and $\tilde{f}_i$ represents the post-collision distribution.

\noindent {\color{red} }

\subsection{Turbulent pipe flow simulation setup}

The computational grid chosen was $n_x \times n_y \times n_z = 300 \times 300 \times 600 $ with the flow driven in the $z$ direction. The pipe radius in lattice units is $R = 149.5$, which results in a streamwise length $L \approx 4.01R $. The flow is driven by a constant body force per unit volume $\rho h$, which, at a fully developed stage should balance the viscous force \cite{Peng2018}, resulting in a friction velocity $u^*$ of
\be
    u^* = \sqrt{\frac{hR}{2}} \ .
\ee 

The viscosity and friction velocity in lattice units are, respectively, $\nu = 0.0032$ and $u^* = 0.00388$, and the friction Reynolds number $Re_{\tau} = u^*R/\nu \approx 181.3$. The viscous length scale is $y^* = \nu/u^*$, which gives the grid spacing in wall units of $\delta_x^+ = \delta_y^+ = 1.212$. All observables with the superscript $+$ from now on, are normalized by the viscous length scale $y^*$ or friction velocity $u^*$.

The no-slip boundary condition is then applied through Eqs.~(\ref{fbc1}) and (\ref{fbc2}) and a periodic boundary condition is applied in the streamwise direction. The large eddy turnover time in lattice units is $T^{LET} =R/u^* = 38531$ and the Reynolds number based on the bulk velocity and on the pipe's diameter is $Re \approx 5329.3 $.

Previous works have already shown that for $Re_{\tau} = 180$ the smallest length scale in turbulent pipe flow is $\eta^+ = 1.5$ \cite{Peng2018}, which is higher than the aforementioned grid spacing in the pipe's cross-section. The pipe length used $L\approx 4.01R$, is compared with different datasets also with periodic boundary conditions and with $L\approx (10R, 12.11R)$ \cite{Loulou1997,Peng2018}. Although it does not meet the suggested length to avoid the effects of periodic boundary conditions on turbulent statistics \cite{Chin2010}, the current pipe length was inspired in recent experimental findings which suggests that structures in turbulent pipe flow are correlated up to $L \approx 4R$ for $Re = 24400$ \cite{JackelPRF2023}.

The flow was initialized with the mean profile
\be
    U(\delta^+) = \begin{cases} \delta^+, & \mbox{if } \delta^+ \leq 10.8 \ , \\ \frac{1}{0.4}ln(\delta^+) + 5.0 \ , & \mbox{if } \delta^+ > 10.8 \ , \end{cases}
\ee

\noindent where $\delta = R - r$ is the pipe's wall distance. To trigger turbulence, in addition to the constant body force, it was added a non-uniform and divergence-free force during the first three $T^{LET}$ \cite{Peng2018}, which is given by
\bea
    && F_r = -h\kappa A_0 \frac{R}{r}\frac{k_z l }{L}sin\left( \frac{2\pi t}{T} \right) \left \{ 1-cos\left [ \frac{2 \pi (R - r - l_0)}{l} \right ] \right \}cos \left ( k_z \frac{2 \pi z}{L} \right ) cos(k_{\theta}\theta) \ , \nonumber \\
    && F_{\theta} = h(1 - \kappa) A_0 \frac{k_z}{k_{\theta}}\frac{2\pi R}{L}sin\left( \frac{2\pi t}{T} \right) sin\left [ \frac{2 \pi (R - r - l_0)}{l} \right ] cos \left ( k_z \frac{2 \pi z}{L} \right ) sin(k_{\theta}\theta) \ , \nonumber \\
    && F_z = -hA_0 \frac{R}{r}sin\left( \frac{2\pi t}{T} \right) sin\left [ \frac{2 \pi (R - r - l_0)}{l} \right ] sin \left ( k_z \frac{2 \pi z}{L} \right ) cos(k_{\theta}\theta) \ , 
    \label{TurbForce}
\eea 

\noindent where $k_z = 3$ and $k_{\theta} = 2$ are the streamwise and azimuthal wavenumbers of the perturbation force, $T \approx 0.052T^{LET}$ and $A_0 = 50$ are the forcing period and amplitude, respectively, and $\kappa = 0.5$ is the weighting parameter that distributes the force in radial and azimuthal directions. The force to trigger turbulence was applied only in the region $R - l_0 - l \leq r \leq R - l_0$ with $l_0 = 0.2R$ and $l = 0.4R$. The numerical simulation was performed for $30T^{LET}$ with statistics being done with the last $10T^{LET}$, taking the instantaneous vector fields at every $100$ iteration.

\subsection{\label{CS Identification}Dimensionality reduction for coherent structure identification}

The identification of CS used in this paper follows the dimensionality reduction approach which was used for moderate and highly turbulent flow ($5300 \leq Re \leq 35000$) \cite{JackelPOF2023,dennis2014}. The procedure is based on the streamwise velocity-velocity correlation function $R_{uu}$, which uses the fluctuating part of the streamwise component $\delta u_z$, which will be represented by $u_z$ for simplicity. In this case, the correlation function is represented by
\be 
    R_{uu}(r_0 + \Delta r,\Delta \theta) = \frac{\langle u_z(r_0,\theta_0) u_z(r_0 + \Delta r, \theta_0 + \Delta \theta) \rangle_{\theta_0}}{u^2_{z,rms}} \ ,
    \label{newcorrelation}
\ee 

\noindent where $u_{z,rms}^2 = \langle u_z(r_0,\theta_0)^2 \rangle_{\theta_0}$. After the computation of Eq.~(\ref{newcorrelation}), the azimuthally dominant mode is selected as the highest peak in the power of the fast Fourier transform (FFT) of the correlation function from a reference radius $r_0 = 0.78R$.

\section{\label{sim_res_sec} Simulation Results}

\subsection{Instantaneous Observations}

\begin{figure}[h!]
    \centering
    \includegraphics[width=1.00\linewidth]{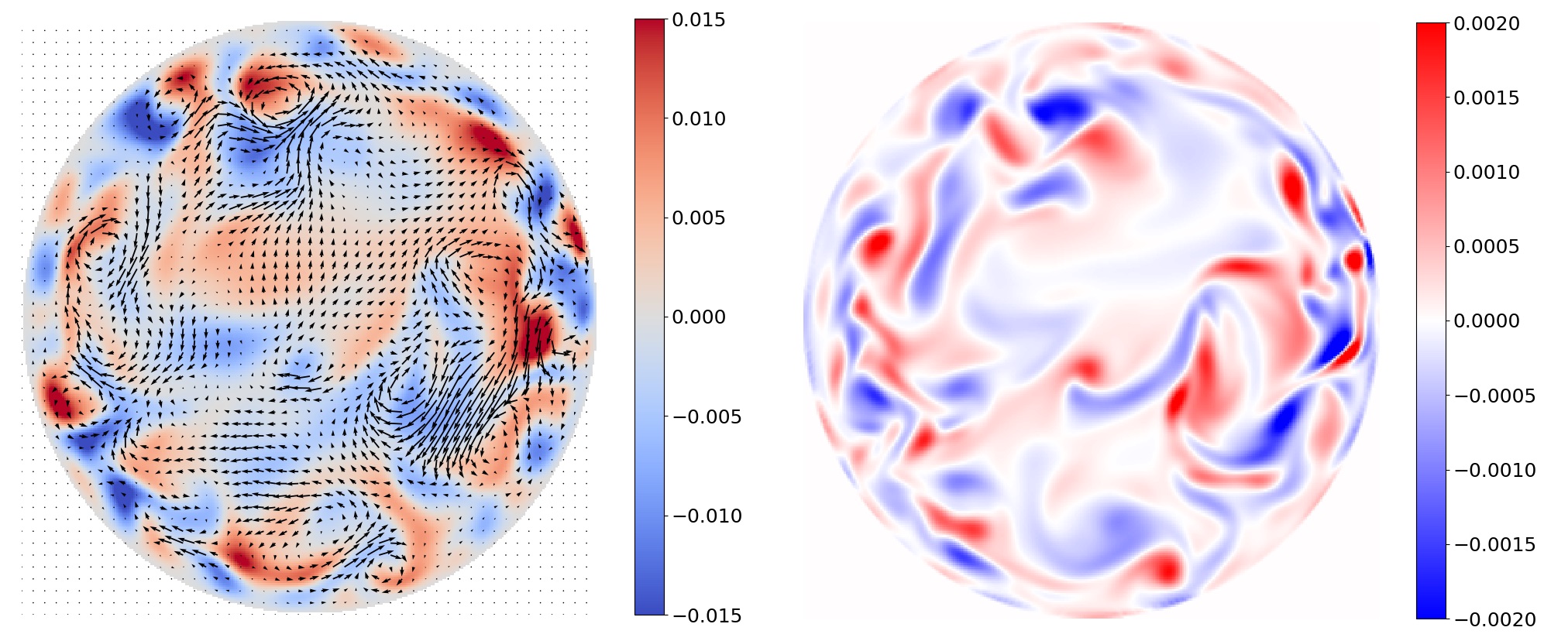}
    \caption{Snapshot observations from turbulent pipe flow simulated through the LBM with $Re_{\tau} \approx 181.3$. Left:In-plane vector field and density plot of streamwise velocity fluctuation in lattice units. Right: Streamwise vorticiy in lattice units.}
    \label{snapshots}
\end{figure}

Instantaneous observations of the turbulent flow are shown in Fig. \ref{snapshots}. As can be seen, the presence of several vortices is displayed randomly distributed through the vector field and also by the streamwise vorticity. It is worth mentioning how the vortices usually appear in regions close to a change of sign in the streamwise velocity fluctuation, in consonance with the well-known picture of ejections and sweeps related to the $Q_2$ and $Q_4$ quadrant analysis \cite{Wallace2016}.

\begin{figure}[!htb]
    \centering
    \includegraphics[width=0.60\linewidth]{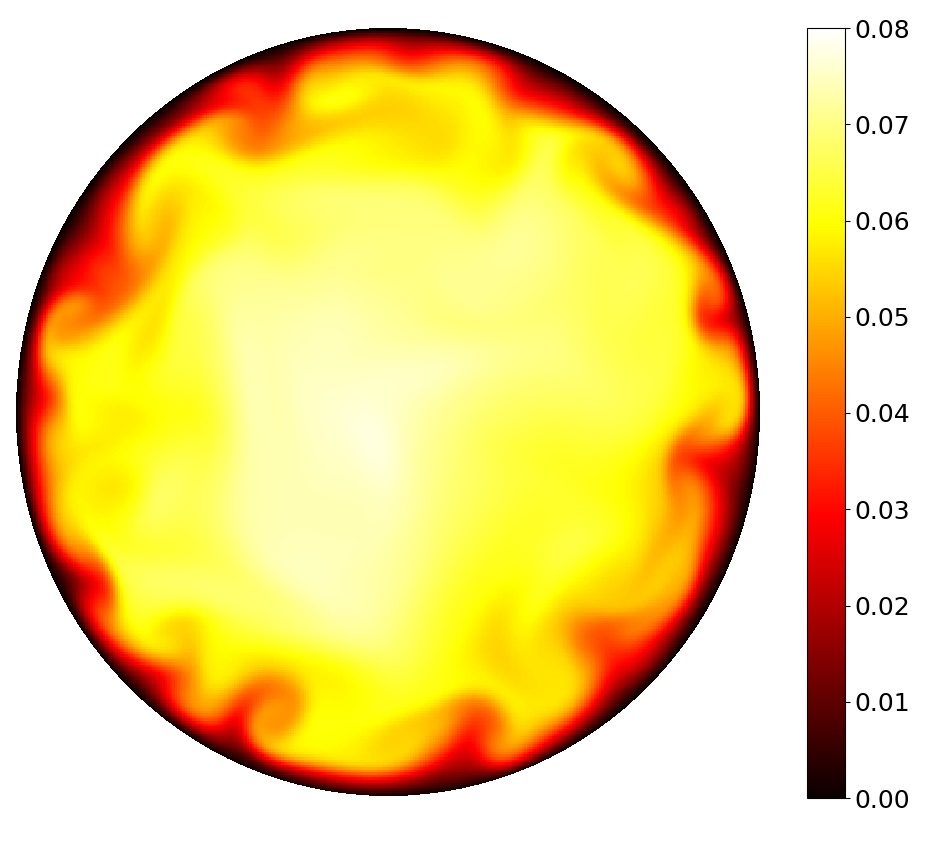}
    \caption{Snapshot of streamwise velocity component $u_z$ in lattice units.}
    \label{streamwiseuz}
\end{figure}

The instantaneous streamwise velocity component $u_z$ is shown in Fig. \ref{streamwiseuz}, where several near-wall structures can be seen, such as the shape of Kelvin-Helmholtz and Rayleigh-Taylor like instabilities. Also, the thickness of the structures seems to agree relatively well with other simulation techniques at $Re_{\tau} = 180$ \cite{Yao2023}, such as the pseudospectral based solver Openpipeflow \cite{Willis2017}.

\begin{figure}[!htb]
    \centering
    \includegraphics[width=0.60\linewidth]{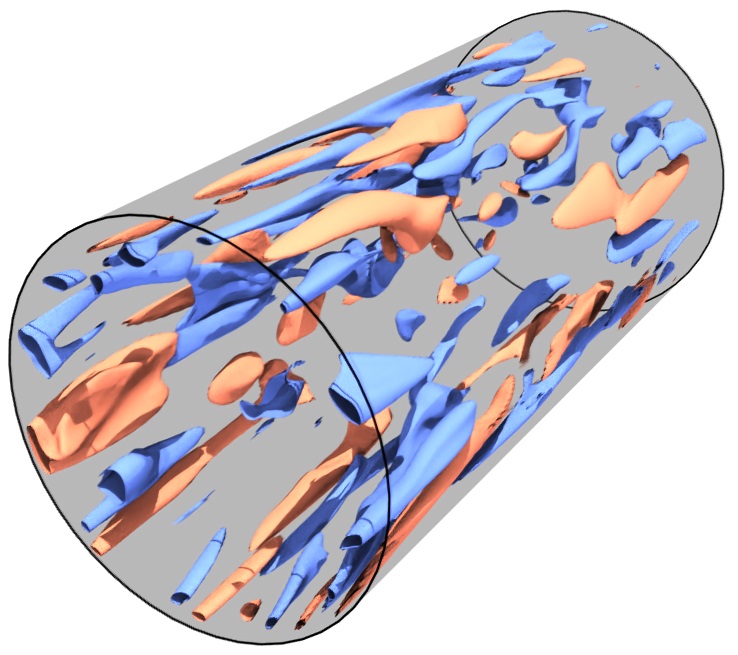}
    \caption{Isosurfaces of streamwise velocity fluctuations. Red/blue isosurfaces are related to high/Low speed streaks, where the velocity fluctuation is $10\%$ above/bellow the mean velocity profile.}
    \label{highandlowspeedstreaks}
\end{figure}

The high/low-speed streaks can be seen by the red/blue isosurfaces of the streamwise velocity fluctuation $\delta u_z$ in Fig. \ref{highandlowspeedstreaks}. Different structures could be targeted by different thresholds on the isosurface selection, decreasing in quantity as the value is increased as it would be related to more extreme events.

\subsection{Turbulent Statistics}

In order to validate the numerical simulations, the turbulent statistics will be compared with a pseudo-spectral approach \cite{Loulou1997}, a LBM approach based on a second order expansion of Maxwell Boltzmann equilibrium distribution --which will be referred to as LBM-$\mathcal{O}(2)$-- \cite{Peng2018}, and with experimental results using Laser Doppler Anemometry (LDA) \cite{Tahitu_thesis}, which is known for its accuracy for near-wall measurements.
\begin{table}[h!]
\centering
\begin{tabular}{|c c c c c|} 
 \hline
 \hline
  & Method & $Re_{\tau}$ & $L/R$ & $\Delta T/T^{LET}$ \\ [0.5ex] 
 \hline
 Peng et al. & LBM-$\mathcal{O}(2)$ & 180.0 & 12.12 & 60.1 \\ 
 \hline
 Loulou et al. & PS & 190.0 & 10.00 & 5.8 \\
 \hline
 Tahitu & LDA & 181.2 & 437.81 & 265.0 \\
 \hline
 Present & LBM-$\mathcal{O}(6)$ & 181.3 & 4.01 & 10.0 \\
 \hline
 \hline
\end{tabular}
\caption{Relevant parameters for turbulent pipe flow' statistical properties.}
\label{table:3}
\end{table}

The relevant parameters for all sets to be used are shown in table \ref{table:3}, where $\Delta T$ is the time window used for the statistical averages and LBM-$\mathcal{O}(6)$ is referred to the scheme 

\begin{figure}[h!]
\centering
\begin{minipage}{.5\textwidth}
  \centering
  \includegraphics[width=1.00\linewidth]{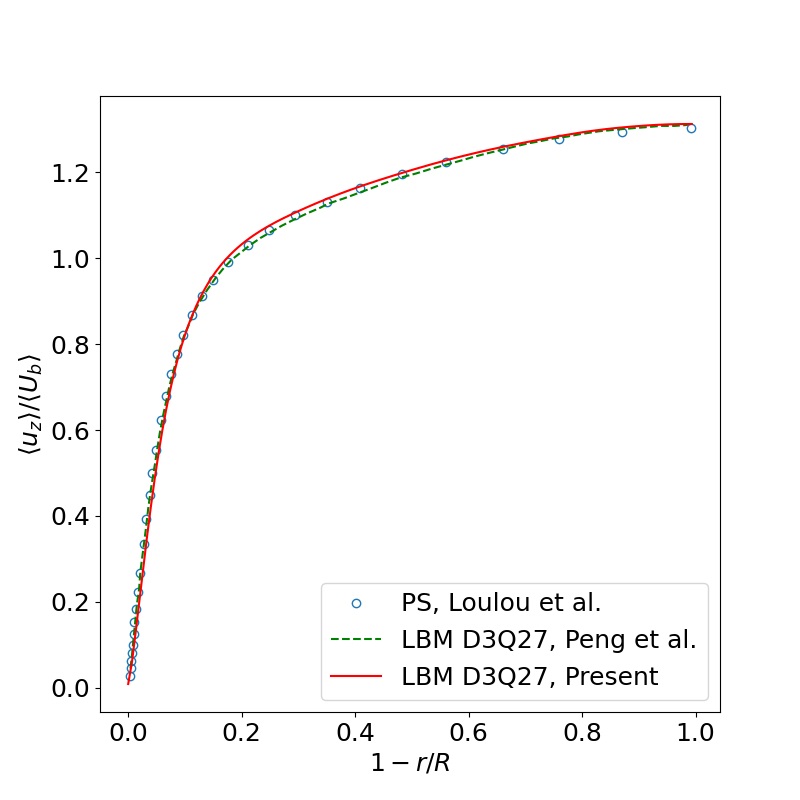}
  \label{meanLBM}
\end{minipage}%
\begin{minipage}{.5\textwidth}
  \centering
  \includegraphics[width=1.00\linewidth]{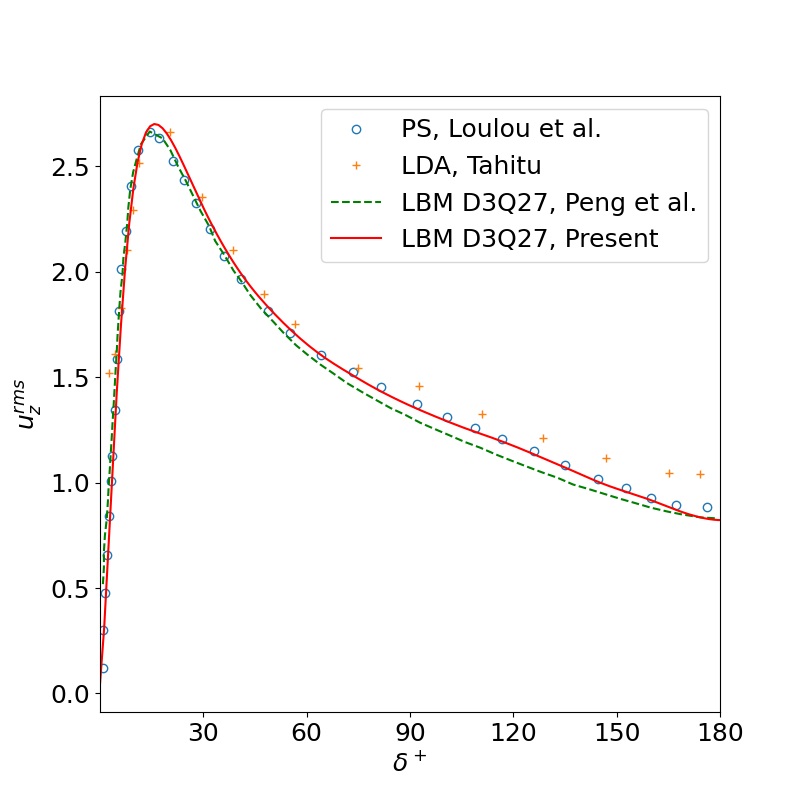}
  \label{urmsLBM}
\end{minipage}
\caption{Left: Mean streamwise velocity profile normalized by the bulk velocity. Right: Root-mean-square streamwise velocity fluctuation in wall units.}
\label{MeanUrms}
\end{figure}

\noindent used in this work with the equilibrium distribution expanded up to the sixth order in Hermite polynomials. The rms of the streamwise velocity fluctuation and high order statistical moments are computed through 
\be
    \mbf{u}_{rms}(\mbf{x}) \equiv \sqrt{\langle \delta\mbf{u}(\mbf{x},t)^2 \rangle_t} \ , \ \  S_n(\mbf{x}) \equiv \frac{\langle \delta \mbf{u}(\mbf{x},t)^n \rangle_t}{\mbf{u}_{rms}(\mbf{x})^n} \ .
    \label{urmsformula}
\ee

\begin{figure}[h!]
\centering
\begin{minipage}{.5\textwidth}
  \centering
  \includegraphics[width=1.00\linewidth]{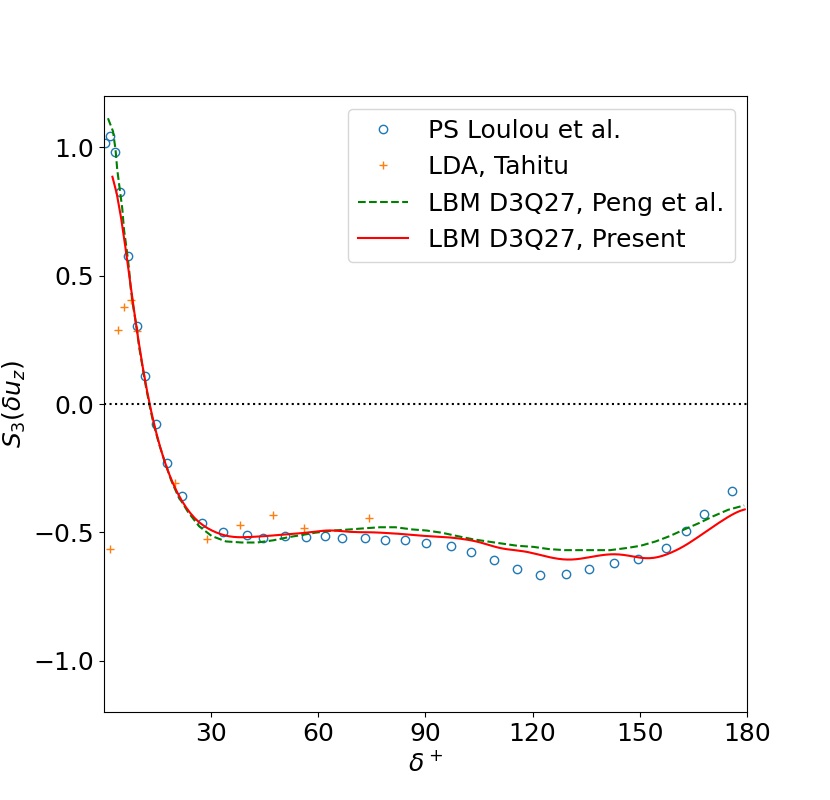}
  \label{skewnessLBM}
\end{minipage}%
\begin{minipage}{.5\textwidth}
  \centering
  \includegraphics[width=1.00\linewidth]{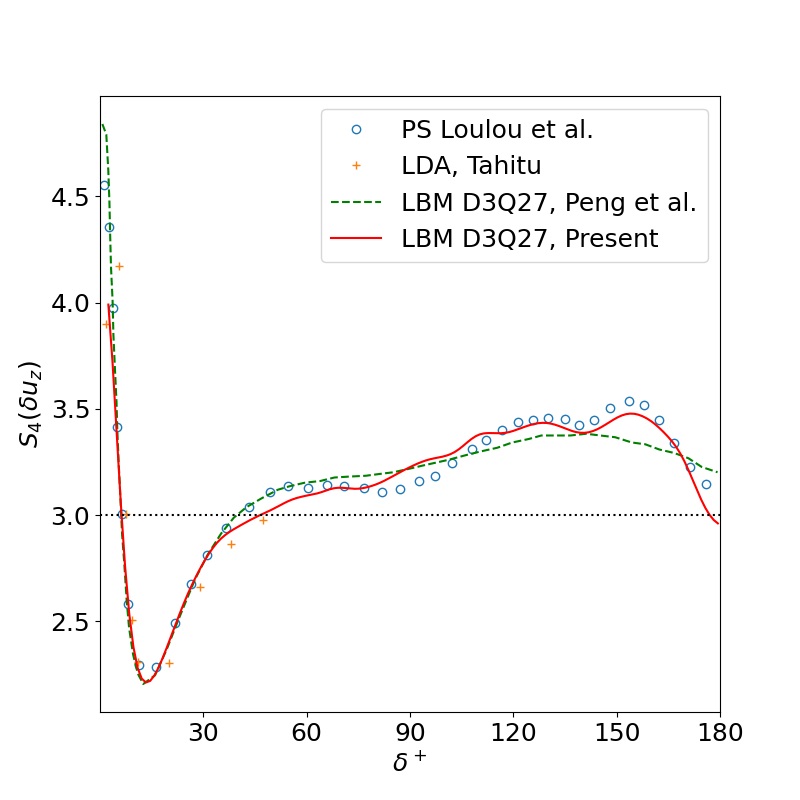}
  \label{flatnessLBM}
\end{minipage}
\caption{High order statistical moments of the streamwise velocity component. Left: Skewness. Right: Flatness.}
\label{S3S4LBM}
\end{figure}

The mean streamwise velocity profile normalized by the bulk velocity and the rms streamwise velocity fluctuation in wall units are shown in Fig. \ref{MeanUrms}. To stress the accuracy of the statistical properties of the present numerical study, skewness and flatness are shown in Fig. \ref{S3S4LBM}, as higher statistical moments require a well converged and large enough statistical ensemble.

As one can see, all observables agree really well with the literature. Although our pipe length is approximately 2.5$\times$ smaller than the PS results, the time window used was almost twice the size in large eddy turnover times. The comparisons with LDA which used a very large time window for the statistical analysis, as one can see in table \ref{table:3}, is also in very good agreement, with the peak positions and plateau of all observables being well represented by the numerical simulation.

\subsection{CS Identification and Dynamics of Mode Transition}

The dimensionality reduction based on the dominant azimuthal wave number of Eq.~(\ref{newcorrelation}) was applied on the simulated turbulent data. The turbulent data set of 10$T^{LET}$ was analyzed at every 20 time steps.

\begin{figure}[h!]
    \centering
    \includegraphics[width=0.60\linewidth]{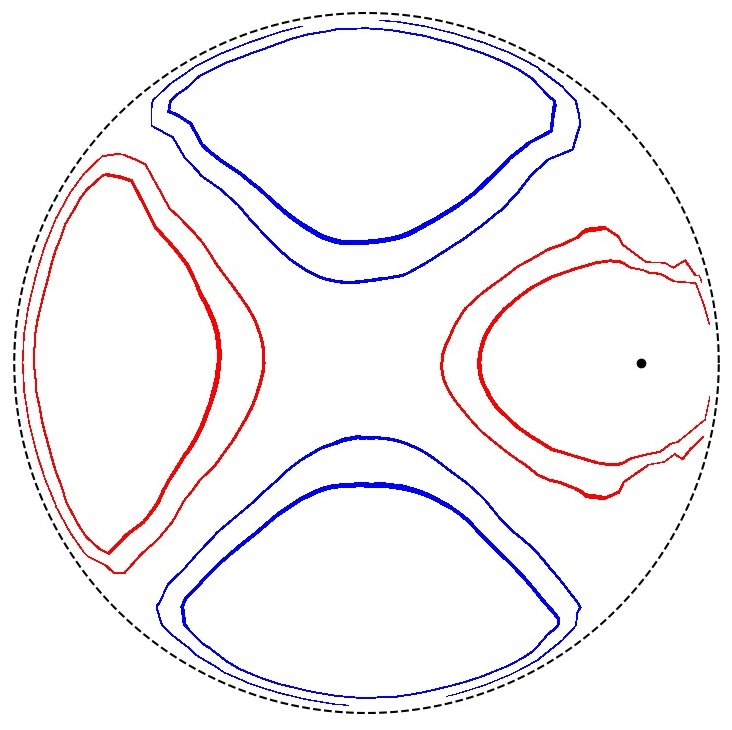}
    \caption{Velocity-velocity correlation contours conditioned to azimuthal wave number $k_{\theta} = 2$. Red countors are related to $R_{uu} = 0.05$ and $0.1$, while blue has the same absolute value with opposite sign.}
    \label{uucorrelationLBM}
\end{figure}

\begin{figure}[h!]
    \centering
    \includegraphics[width=0.80\linewidth]{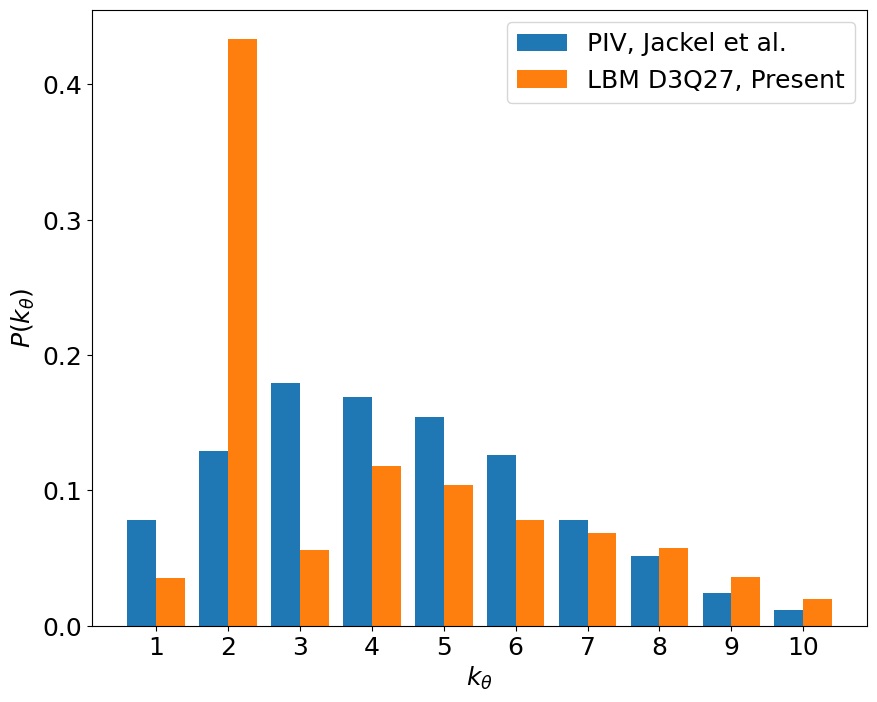}
    \caption{Probability distribution of dominant azimuthal wave numbers.}
    \label{modedistributionLBM}
\end{figure}

Figure \ref{uucorrelationLBM} shows the contour plot of the velocity-velocity correlation function conditioned to the azimuthal wave number $k_{\theta} = 2$. Remarkably, there were found approximately 10 different azimuthal wave numbers in the produced numerical data. Modes higher than mode 10 were also found, but their statistical weight is negligible, as pointed out in the experimental findings \cite{JackelPOF2023,JackelICHMT2023,JackelPRF2023}. The comparison between the current probability distribution of dominant azimuthal wave numbers with the experimental one is shown in Fig. \ref{modedistributionLBM}.

The peak in the azimuthal mode $k_{\theta} = 2$ seems to be related to the \emph{principle of permanence of large eddies} \cite{Frisch} and to a non-trivial memory effect regarding the force that was used to trigger turbulence (\ref{TurbForce}), as the force was only active in the first $3T^{LET}$ and the dimensionality reduction was only applied in the range $20 T^{LET}\leq t \leq 30 T^{LET}$. At the same time, the mode distributions, besides the peaked forced mode, resemble the cascade effects in three-dimensional turbulence, as expected, with the energy going from large to small scales.

\begin{figure}[h!]
    \centering
    \includegraphics[width=1.00\linewidth]{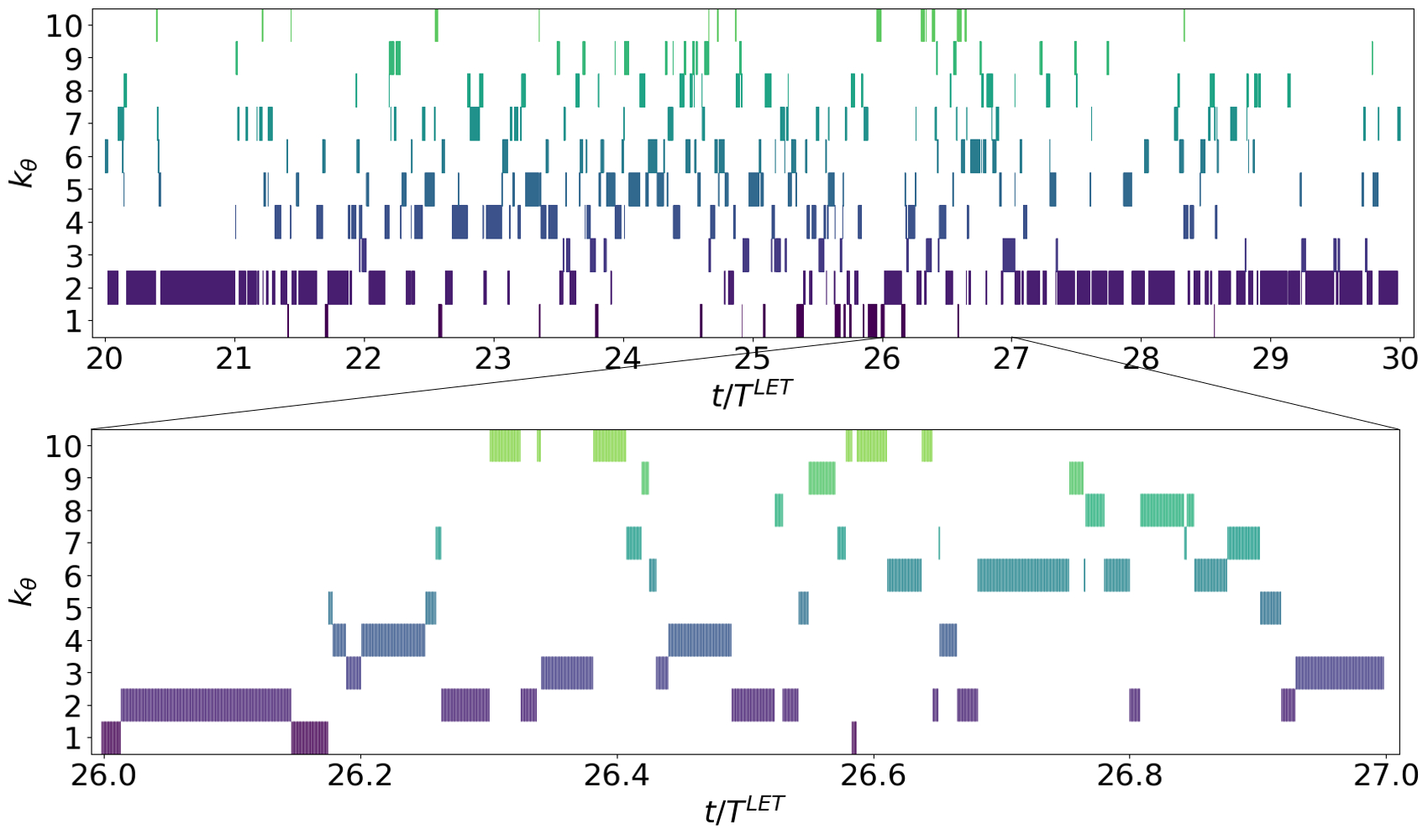}
    \caption{Time dependence of the identified dominant modes.}
    \label{modetransitions}
\end{figure}

The time dependence of the identified modes during the 10 analyzed turnover times is shown in Fig. \ref{modetransitions}. As one can see, the dynamics between the modes seems to be approximately random for a big time interval, however, the intriguing phenomenon is the persistence of mode $k_{\theta} = 2$ (the forced mode) intermittently. To illustrate how the modes are finely resolved in time, a zoom-in of 10$\times$ is also displayed.

\begin{figure}[h!]
    \centering
    \includegraphics[width=0.80\linewidth]{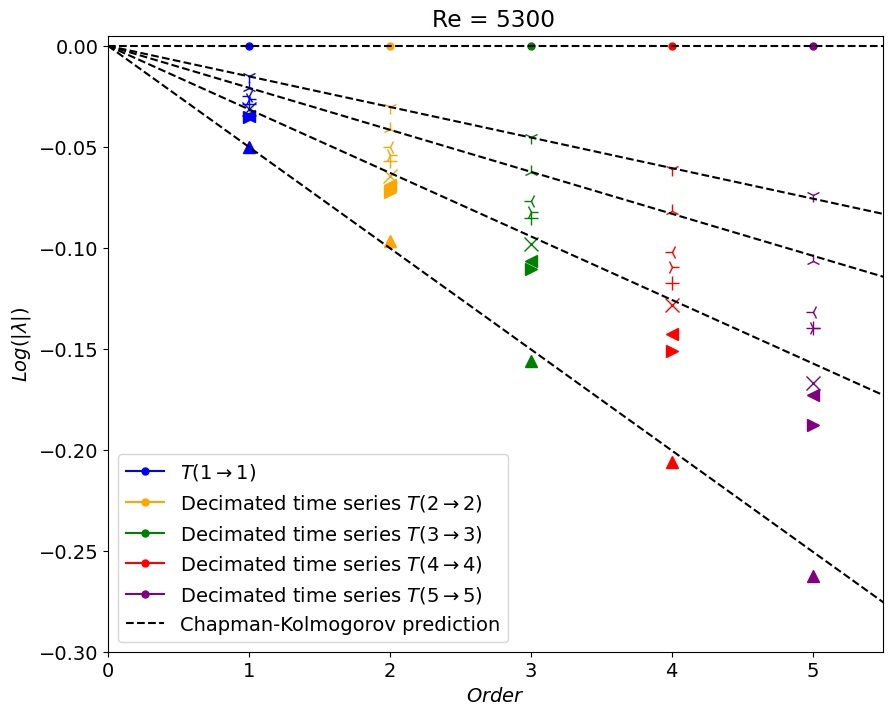}
    \caption{Comparison between eigenvalues of the original transition probability matrix (Order = $1$) with eigenvalues for the transition matrices of the decimated series (Order up to $5$).}
    \label{CKLBM}
\end{figure}

A stochastics series based on the identified modes to investigate its transitions was also set, as in Ref \cite{JackelPRF2023}. In this case, it is given by
\be
    \mathcal{S}^{LBM} \equiv  \{ k^*(t_0), k^*(t_0 + \Delta), \ ... , k^*(30T^{LET}) \  \} \ ,
    \label{stochasticseriesLBM}
\ee

\noindent where $t_0 = 20T^{LET}$ and $\Delta = 20 \delta t$. The transition matrix among the identified modes is given by
\be
    T = 
    \begin{bmatrix}
    0.975 & 0.000 & 0.001 & 0.000 & 0.002 & 0.001 & 0.001 & 0.003 & 0.000 & 0.011\\
    0.004 & 0.992 & 0.006 & 0.004 & 0.006 & 0.008 & 0.008 & 0.009 & 0.004 & 0.011\\
    0.003 & 0.001 & 0.977 & 0.002 & 0.001 & 0.003 & 0.002 & 0.001 & 0.000 & 0.005\\
    0.000 & 0.002 & 0.007 & 0.980 & 0.002 & 0.003 & 0.002 & 0.002 & 0.006 & 0.005\\
    0.003 & 0.001 & 0.004 & 0.002 & 0.978 & 0.003 & 0.005 & 0.005 & 0.004 & 0.003\\
    0.001 & 0.001 & 0.001 & 0.003 & 0.005 & 0.973 & 0.003 & 0.006 & 0.004 & 0.003\\
    0.004 & 0.001 & 0.004 & 0.002 & 0.005 & 0.002 & 0.969 & 0.003 & 0.004 & 0.005\\
    0.006 & 0.001 & 0.000 & 0.002 & 0.002 & 0.002 & 0.005 & 0.969 & 0.004 & 0.003\\
    0.000 & 0.000 & 0.000 & 0.002 & 0.001 & 0.003 & 0.004 & 0.003 & 0.970 & 0.000\\
    0.003 & 0.000 & 0.001 & 0.001 & 0.001 & 0.002 & 0.002 & 0.000 & 0.001 & 0.953
    \end{bmatrix} \ , \
    \label{transitionmatrixlbm}
\ee 

\noindent which, as one can see, is diagonally dominant, with $k_{\theta} = 2$ being the mode with a higher probability of persistence. The Chapman-Kolmogorov (CK) equation was also tested by performing a decimation of several orders on the stochastic series (\ref{stochasticseriesLBM}). The results of the CK test are shown in Fig. \ref{CKLBM}. In contrast with the experimental case reported in Ref \cite{JackelPRF2023}, a qualitatively good agreement was found between the comparison of the absolute values of the eigenvalues from the original series with their respective counterparts from the decimated series.

This was observed up to a fifth-order decimation, which reduces the amount of analyzed turbulent fields by a factor of $1/100$, resulting in approximately 3850 snapshots, which might not be large enough, so the effects of low statistics may play a role. Also, the deviation from the CK prediction could be related to a transition to the scenario observed in the experiment reported in Ref \cite{JackelPRF2023}, where the dynamics among the transitions display a non-Markovian behavior. It is worth mentioning that the temporal resolution obtained by the LBM is of the order of $10^2$ finer than the experimental case obtained with a laser of 15 Hz. In this case, the agreement with the CK prediction could be associated with the fact that the transition matrix is diagonally dominant, so the modes, in general, don't have a considerably high memory about the transition from different modes.

\begin{figure}[!htb]
    \centering
    \includegraphics[width=0.75\linewidth]{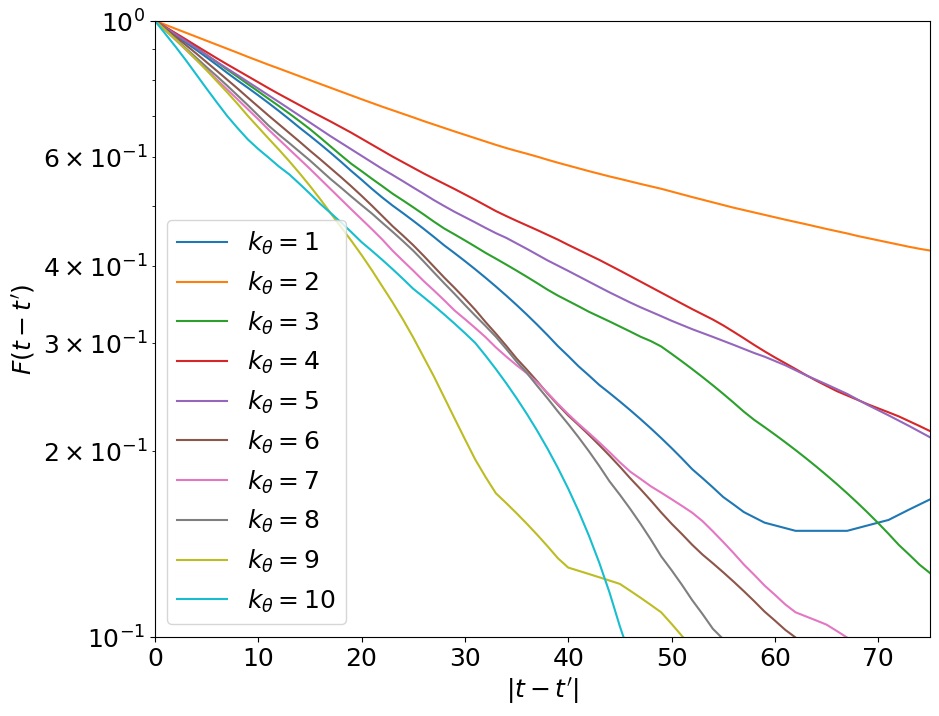}
    \caption{Semi-log plot of self mode correlation for all the identified modes.}
    \label{FcorrelationLBM}
\end{figure}

\begin{figure}[!htb]
    \centering
    \includegraphics[width=0.75\linewidth]{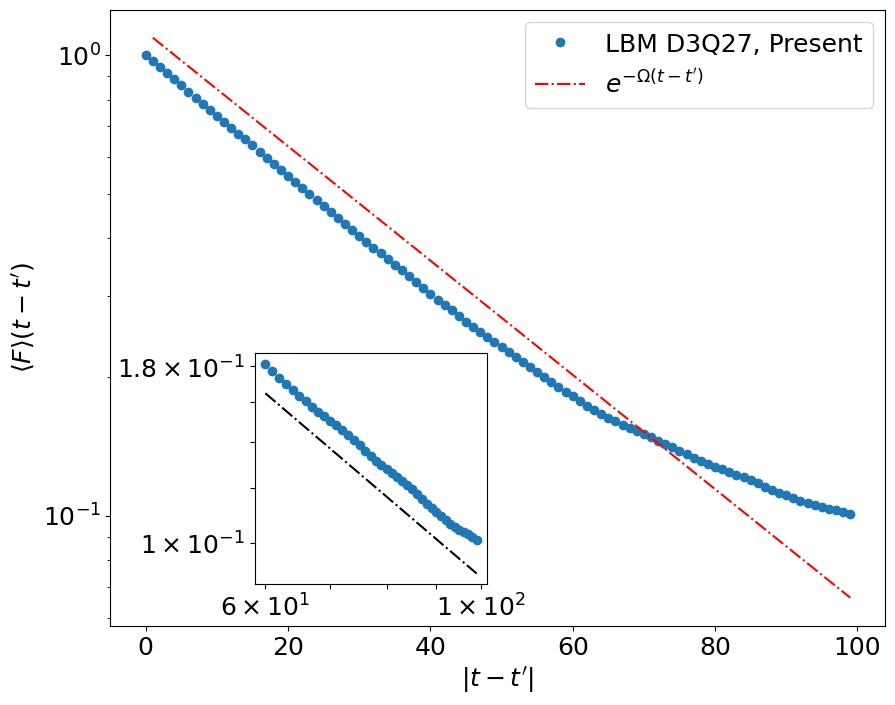}
    \caption{Semi-log plot of average self correlation and straight line of exponential decay for comparison. Inset shows the behavior of region $\vert t - t'\vert \geq 60$ in a log-log plot with a straight line of algebric decay with exponent $-1.2$ for comparison.}
    \label{FmeancorrelationLBM}
\end{figure}

To investigate quantitatively the observed agreement between the CK prediction with the decimated stochastic time series, it was analyzes the self mode correlation function \cite{JackelPRF2023}
\be
    \Tilde{F}(t-t') \equiv \langle V_m(t) \cdot V_m(t') \rangle -  \langle V_m \rangle ^2 \ , \ F(t-t') \equiv \frac{\Tilde{F}(t-t')}{\Tilde{F}(0)} \ , \  
\ee

\noindent were $V_m(t) = \delta_{m,k^*(t)}$ and $ 1\leq m \leq 10$. Firstly the correlation function was investigated individually, for each mode, and afterwards mode averaged, which will be represented by $\langle F \rangle (t - t')$. Figure \ref{FcorrelationLBM} shows that, in fact, the correlation function for all modes decays approximately exponentially, which is observed as a straight line in the semi-log plot. Also, as one can see, mode $k_{\theta} = 2$ is the most self correlated one, which agrees with the observed probability distribution. The average self correlation is displayed in Fig. \ref{FmeancorrelationLBM} together with a straight line representing an exponential decay with $\Omega = 1/35$, as it can be seen, the agreement is really evident and it holds till $\vert t - t'\vert \approx 60$. Larger time separations are shown in the inset, in a log-log plot, which was observed to decay algebraically $\propto \vert t - t'\vert^{-\alpha}$, as indicated by the black straight line, with $\alpha = 1.2$. This fact agrees with previous experimental findings of non-Markovian behavior in turbulent pipe flow with $\alpha \approx 1.0$ for $Re = 24400$, recorded with a smaller frequency \cite{JackelPRF2023}.

The largest and mean streamwise lengths for all modes are presented in table \ref{table:4} by means of Taylor frozen hypothesis \cite{Taylor1938}.
\begin{table}[h!]
\centering
\begin{tabular}{|c c c c c c c c c c c|} 
 \hline
 \hline
 $k_{\theta}$ & 1 & 2 & 3 & 4 & 5 & 6 & 7 & 8 & 9 & 10 \\ [0.5ex] 
 \hline
 $\Delta S_{max}/R$ & 0.98 & 8.20 & 1.33 & 1.73 & 1.50 & 1.01 & 1.01 & 0.70 & 0.50 & 0.49 \\ 
 \hline
 $\langle\Delta S \rangle/R$ & 0.27 & 0.87 & 0.32 & 0.36 & 0.33 & 0.27 & 0.21 & 0.23 & 0.22 & 0.15 \\ 
 \hline
 \hline
\end{tabular}
\caption{The largest and average sizes of the coherent structure mode lengths from the Taylor frozen hypothesis.}
\label{table:4}
\end{table}
As one can see, on average, all modes are within the simulated pipe's length. Although, their longest example might by related to the known LSM and VLSM --which are typically longer than three pipe radii \cite{adrian2007,dennis2014,Balakumar2007}.

\section{\label{conc_sec} CONCLUSIONS}

The turbulent pipe flow simulation was performed by the LBM and validated extensively against experimental and different numerical methodologies. The turbulent data is resolved in space up to the viscous length scale and the instantaneous observations reveal the seemingly random distribution of CSs. Our pipe length followed the suggestion of the size of correlated structures in experimental turbulent pipe flow, as determined in Ref. \cite{JackelPRF2023}, and the statistical moments of streamwise velocity fluctuation matched well previous results in the literature. The CS detection was performed in the same fashion as in previous experimental cases \cite{JackelPRF2023,JackelPOF2023,dennis2014,JackelICHMT2023}, which revealed a non-trivial correlation with the force used to trigger the turbulent state. Also, the mode distribution agrees with the experimental findings regarding the turbulence cascade of energy from large to small scales.

The stochastic mode transitions for the finely time-resolved data resulted in a diagonally dominant transition matrix, which culminated in a good agreement of the CK prediction for the decimated time series. The indicated Markovian behavior was then checked by time self-mode correlation functions, which confirmed the exponential decay for all observed modes in a short time window. The average self-mode correlation was plotted for a larger time separation, which revealed a transition from the exponential decay to an algebraically one, recovering the observed experimental behavior of non-Markovianity. The streamwise length of the modes by Taylor's frozen hypothesis showed that, on average, they are all within the length of our pipe simulations, but their largest instances might be related to the known LSM.

This work opens the path for many directions of further research. From the experimental point of view, it would be interesting to investigate the CS detection and stochastic mode transitions for higher frequencies with larger data sets, since this would allow the observation of the Markovian effect for small-time separations, transitioning to a non-Markovian behavior (as already observed), and finally transitioning back to a Markovian behavior for larger time separations when the memory effects of the structure decay exponentially. From the numerical point of view, it would be interesting to investigate larger time series, so that the decimation could go up to a higher order. A 20x higher should be enough to see the same effects of the experimental case with the reported acquisition frequency \cite{JackelPRF2023}. New numerical simulations forcing all modes evenly or with the statistical distribution of the previous literature reports \cite{JackelPRF2023,JackelPOF2023,dennis2014,JackelICHMT2023, Schneider2007,Hellstrom2017} could give a closer result to the real experiment. Finally, modeling the transition between the identified modes is also valuable information that could describe the effects of the turbulent structures more physically.

\vspace{0.5cm}

\leftline{\it{Acknowledgments}}
\vspace{0.3cm}

The authors gratefully acknowledge the financial support from the Conselho Nacional de Desenvolvimento Científico e Tecnológico (CNPq), Brazil and Shell Brazil through the project “Predictive modelling of inorganic scaling in pipes and equipment subjected to multiphase flows in Pre-Salt conditions” at the Interdisciplinary Center of Fluid Dynamics (NIDF) and the strategic importance of the support given by ANP through the R\&D levy regulation.

\nocite{*}

\vspace{0.5cm}

\leftline{\it{Data Availability}}
\vspace{0.3cm}

The data that support the findings of this study are available from the corresponding author upon reasonable request.

\appendix

{\parindent0pt
\section{\small{GENERAL MRT MODEL WITH EQUILIBRIUM DISTRIBUTION EXPANDED UP TO THE SIXTH ORDER IN HERMITE POLYNOMIALS}}
\label{Appendix}
}

As this work is aimed to perform three-dimensional pipe flow simulations, we restrict to the case of a D3Q27 lattice model, which is represented by the lattice velocities $\bs{c}_i = (\vert c_{ix} \rangle,\vert c_{iy} \rangle,\vert c_{iz} \rangle)$ as
%\begin{widetext}
\bea
    && \vert c_{ix} \rangle =  (0, 1,-1, 0, 0,0 0, 1,-1, 1,-1, 1,-1, 1,-1, 0, 0,0, 0, 1,-1, 1,-1, 1,-1, 1,-1)^T  \ , \  \nonumber \\
    && \vert c_{iy} \rangle =  (0, 0, 0, 1,-1, 0, 0, 1, 1,-1,-1, 0, 0, 0, 0, 1,-1, 1,-1, 1, 1,-1,-1, 1, 1,-1,-1)^T \ , \ \nonumber \\
    && \vert c_{iz} \rangle = (0, 0, 0, 0, 0, 1,-1, 0, 0, 0, 0, 1, 1,-1,-1, 1, 1,-1,-1, 1, 1, 1, 1,-1,-1,-1,-1)^T \ . \  \nonumber \\
\eea
%\end{widetext} 

\begin{figure}[h!]
    \centering
    \includegraphics[width=1.00\linewidth]{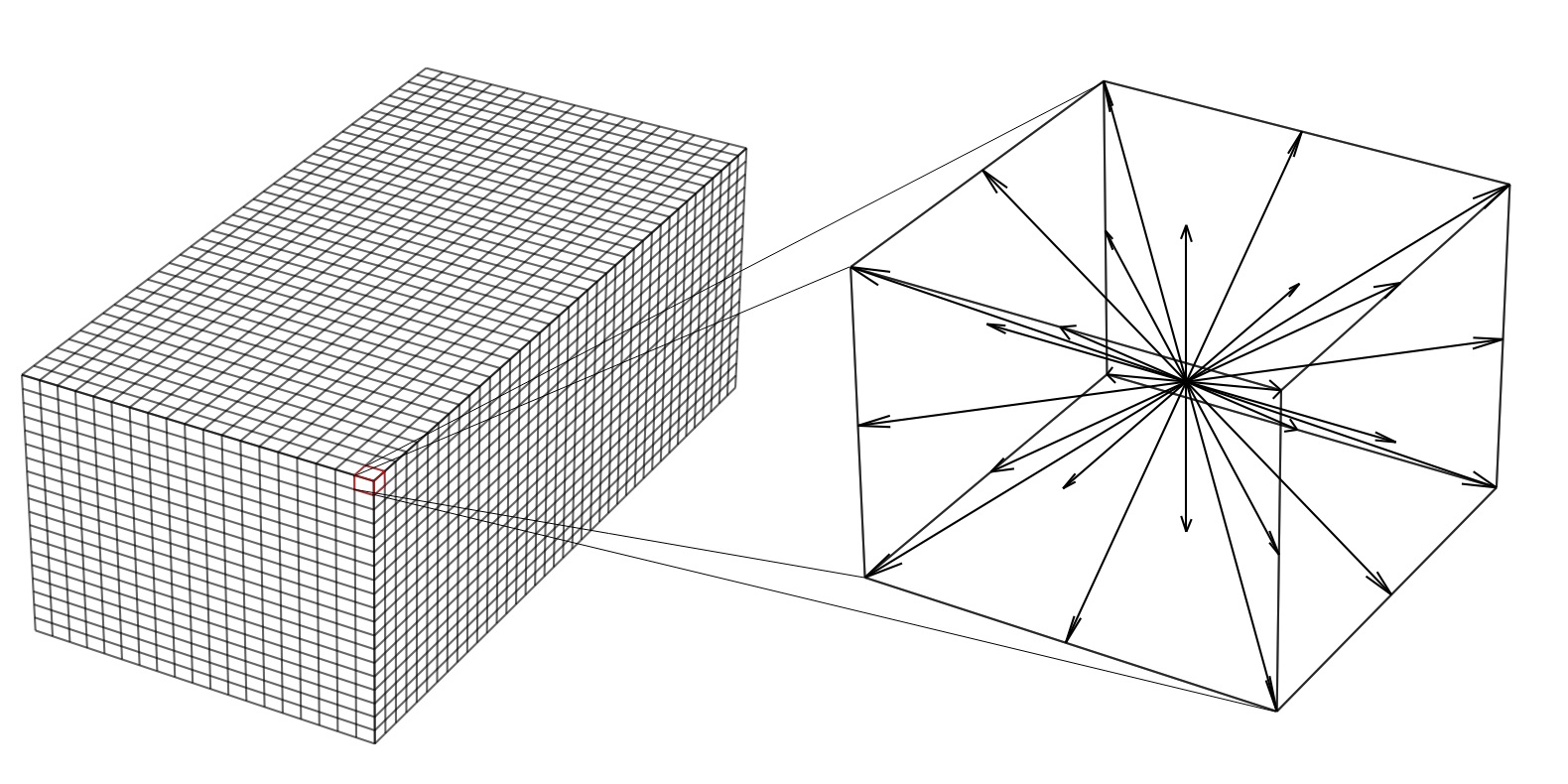}
    \caption{Illustration of computational grid. Each cell is composed by the 27 lattice vectors.}
    \label{gridandcell}
\end{figure}

The computational grid is illustrated in Fig. \ref{gridandcell}. The main idea from the MRT collision model is to transform the usual space of populations,
\be
\vert f_i \rangle = (f_0,...,f_i,...,f_{26})^T \ , \ 
\ee
to the space of velocity moments,
\be
\vert m_i \rangle \equiv (m_0,...,m_i,...,m_{26})^T = \bs{M}^T \vert f_i \rangle  \ , \ 
\ee

\noindent where $\bs{M}$ is the transformation matrix, which is given by \cite{derosis2017},
\begin{eqnarray}
&& \vert M_{0} \rangle = \vert 1,...,1 \rangle \ , \ 
 \vert M_{1} \rangle = \vert c_{ix} \rangle \ , \  
 \vert M_{2} \rangle = \vert c_{iy} \rangle \ , \ 
 \vert M_{3} \rangle = \vert c_{iz} \rangle \ , \ 
\vert M_{4} \rangle = \vert c_{ix}c_{iy} \rangle \ , \  \nonumber \\
&& \vert M_{5} \rangle = \vert c_{ix}c_{iz} \rangle \ , \ 
\vert M_{6} \rangle = \vert c_{iy}c_{iz} \rangle \ , \ 
\vert M_{7} \rangle = \vert c_{ix}^2 - c_{iy}^2 \rangle \ , \ 
\vert M_{8} \rangle = \vert c_{ix}^2 - c_{iz}^2 \rangle \ , \ \nonumber \\
&& \vert M_{9} \rangle = \vert c_{ix}^2 + c_{iy}^2 + c_{iz}^2 \rangle \ , \  \vert M_{10} \rangle = \vert c_{ix} c_{iy}^2 + c_{ix} c_{iz}^2 \rangle \ , \ \vert M_{11} \rangle = \vert c_{ix}^2 c_{iy} + c_{iy} c_{iz}^2 \rangle \ , \  \nonumber \\
&& \vert M_{12} \rangle = \vert c_{ix}^2c_{iz} + c_{iy}^2c_{iz} \rangle \ , \   \vert M_{13} \rangle = \vert c_{ix} c_{iy}^2 - c_{ix} c_{iz}^2 \rangle \ , \  \vert M_{14} \rangle = \vert c_{ix}^2 c_{iy} - c_{iy} c_{iz}^2 \rangle \ , \  \nonumber \\
&& \vert M_{15} \rangle = \vert c_{ix}^2c_{iz} - c_{iy}^2c_{iz} \rangle \ , \ \vert M_{16} \rangle = \vert c_{ix} c_{iy} c_{iz} \rangle \ , \  \vert M_{17} \rangle = \vert c_{ix}^2 c_{iy}^2 + c_{ix}^2 c_{iz}^2 + c_{iy}^2 c_{iz}^2 \rangle \ , \  \nonumber \\
&& \vert M_{18} \rangle = \vert c_{ix}^2 c_{iy}^2 + c_{ix}^2 c_{iz}^2 - c_{iy}^2 c_{iz}^2 \rangle \ , \   \vert M_{19} \rangle = \vert c_{ix}^2 c_{iy}^2 - c_{ix}^2c_{iz}^2 \rangle \ , \  \vert M_{20} \rangle = \vert c_{ix}^2 c_{iy} c_{iz}  \rangle \ , \  \nonumber \\
&& \vert M_{21} \rangle = \vert c_{ix} c_{iy}^2 c_{iz} \rangle \ , \   \vert M_{22} \rangle = \vert c_{ix} c_{iy} c_{iz}^2 \rangle \ , \   \vert M_{23} \rangle = \vert c_{ix} c_{iy}^2 c_{iz}^2  \rangle \ , \  \vert M_{24} \rangle = \vert c_{ix}^2 c_{iy} c_{iz}^2 \rangle \ , \  \nonumber \\
&& \vert M_{25} \rangle = \vert c_{ix}^2 c_{iy}^2 c_{iz} \rangle \ , \  \vert M_{26} \rangle = \vert c_{ix}^2 c_{iy}^2 c_{iz}^2  \rangle \ . \
\label{Mmatrix}
\end{eqnarray}

The MRT collision operator is written as 
\be
    \boldsymbol{L}[\vert f_i \rangle] \equiv -(\bs{M}^T)^{-1}\bs{\Lambda} \bs{M}^T(\vert f_i \rangle - \vert f_i^{eq} \rangle) \ , \  \label{mrt}
\ee

\noindent where the matrix elements of the $27 \times 27$ collision matrix $\mathbf{\Lambda}$ are defined as \cite{Yoshida2010,coveney_succi_dhumieres_ginzburg2002,hosseini2019}
\be
    \Lambda = diag\left [ 1,1,1,1,\omega,\omega,\omega,\omega,\omega,1,...,1 \right ] \ ,
    \label{lambda}
\ee 

\noindent where $\omega = 1/\tau = (\nu/c_s^2 + 0.5)^{-1}$ with $\nu$ being the kinematic viscosity.

Using the MRT collision operator defined in Eq.~(\ref{mrt}), the equation for the post collision distributions (\ref{collision}) now becomes 
\be
    \vert m_i^* \rangle = (\mbf{I} - \mbf{\Lambda})\vert m_i \rangle + \mbf{\Lambda} \vert m_i^{eq} \rangle + \Bigg ( \mbf{I} - \frac{\mbf{\Lambda}}{2} \Bigg )\vert R_i^{MRT} \rangle \ , \ \label{m*}
\ee

\noindent where $\vert R_i^{MRT} \rangle$ is the forcing term within the MRT collision approach, which is given by \cite{guo2002}
\be
\vert R^{MRT}_i \rangle = - \bs{M}^T (\mathbf{F} \cdot \nabla_{\bf{c}}) \vert f_i^{eq} \rangle \ , \ \label{Rmrti}
\ee

\noindent where $\mbf{F} = (F_x,F_y,F_z)$ is an arbitrary external force.

In this work, the Maxwell Boltzmann equilibrium distribution is expanded up to the sixth order in Hermite polynomials $\mathcal{H}_{i}^{(n)}$ since it was already shown that the D3Q27 lattice model can only active its full potential by making use of the 27 Hermite polynomials \cite{derosis_huang_coreixas2019,malaspinas2015,Coreixas_wissocq_puigt_boussuge_sagaut2017,coreixas_phd_thesis,coreixas_chopard_latt2019,derosis2017,derosis_luo2019}. The equilibrium distribution is then written as
%\begin{widetext}
\begin{eqnarray}
            &&f_i^{eq} =  \omega_i \rho \Bigg  \{ 1 + \frac{\mathbf{c}_i \cdot \mathbf{u}}{c_s^2} + \frac{1}{2c_s^4} \Bigg [ \mathcal{H}_{ixx}^{(2)}u_x^2 + \mathcal{H}_{iyy}^{(2)}u_y^2 + \mathcal{H}_{izz}^{(2)}u_z^2 + 2 \Bigg ( \mathcal{H}_{ixy}^{(2)}u_x u_y + \mathcal{H}_{ixz}^{(2)}u_x u_z + \nonumber \\
            &+&\mathcal{H}_{iyz}^{(2)}u_y u_z \Bigg ) \Bigg ] + \frac{1}{2c_s^6} \Bigg [ \mathcal{H}_{ixxy}^{(3)}u_x^2 u_y + \mathcal{H}_{ixxz}^{(3)}u_x^2 u_z + \mathcal{H}_{ixyy}^{(3)}u_x u_y^2 + \mathcal{H}_{ixzz}^{(3)}u_x u_z^2 + \mathcal{H}_{iyzz}^{(3)}u_y u_z^2 +  \nonumber \\
            &+&\mathcal{H}_{iyyz}^{(3)}u_y^2 u_z +2 \mathcal{H}_{ixyz}^{(3)}u_x u_y u_z \Bigg ] + \frac{1}{4c_s^8} \Bigg [ \mathcal{H}_{ixxyy}^{(4)}u_x^2 u_y^2 + \mathcal{H}_{ixxzz}^{(4)}u_x^2 u_z^2 + \mathcal{H}_{iyyzz}^{(4)}u_y^2 u_z^2 + \nonumber \\
            &+&2 \Bigg( \mathcal{H}_{ixyzz}^{(4)}u_x u_y u_z^2 + \mathcal{H}_{ixyyz}^{(4)}u_x u_y^2 u_z + \mathcal{H}_{ixxyz}^{(4)}u_x^2 u_y u_z \Bigg ) \Bigg ] + \frac{1}{4c_s^{10}} \Bigg [ \mathcal{H}_{ixxyzz}^{(5)}u_x^2 u_y u_z^2 + \nonumber \\ 
            &+& \mathcal{H}_{ixxyyz}^{(5)}u_x^2 u_y^2 u_z + \mathcal{H}_{ixyyzz}^{(5)}u_x u_y^2 u_z^2 \Bigg ] + \frac{1}{8c_s^{12}}\mathcal{H}_{ixxyyzz}^{(6)}u_x^2 u_y^2 u_z^2  \Bigg \} \ , \ 
            \label{BoltzmannEquilibriumDistribution}
\end{eqnarray}
where $c_s = 1/\sqrt{3}$ is the lattice sound velocity, and $\omega_i$ are lattice-Boltzmann weights, defined by 
\begin{equation}
\omega_1 =...= \omega_6 = \omega_0/4 \ , \ \omega_7 =...= \omega_{18} = \omega_0/16 \ , \ \omega_{19} =...= \omega_{26} = \omega_0 / 64 \ , \ \label{omegas}
\end{equation}
with $\omega_0 = 8/27$.

Making use of the transformation matrix (\ref{Mmatrix}) and the sixth-order expanded Maxwell Boltzmann equilibrium distribution (\ref{BoltzmannEquilibriumDistribution}), the equilibrium moments can be written as
\bea 
    && m_0^{eq} = \rho \ , \ \ m_1^{eq} = \rho u_x \ , \ \ m_2^{eq} = \rho u_y \ , \ \ m_3^{eq} = \rho u_z \ , \ \ m_4^{eq} = \rho u_xu_y \ ,  \nonumber \\
    && m_5^{eq} = \rho u_xu_z \ , \ \ m_6^{eq} = \rho u_yu_z \ , \ \ m_7^{eq} = \rho (u_x^2 - u_y^2) \ , \ \ m_8^{eq} = \rho (u_x^2 - u_z^2) \ , \nonumber \\
    && m_9^{eq} = \rho (u_x^2 + u_y^2 + u_z^2 + 1) \ , \ \ m_{10}^{eq} = \rho c_s^2 u_x (3u_y^2 + 3u_z^2 + 2) \ , \nonumber \\
    && m_{11}^{eq} = \rho c_s^2 u_y (3u_x^2 + 3u_z^2 + 2) \ , \ \  m_{12}^{eq} = \rho c_s^2 u_z (3u_x^2 + 3u_y^2 + 2) \ , \nonumber \\
    && m_{13}^{eq} = \rho u_x (u_y^2 - u_z^2) \ , \ \ m_{14}^{eq} = \rho u_y (u_x^2 - u_z^2) \ , \ \ m_{15}^{eq} = \rho u_z (u_x^2 - u_y^2) \ , \nonumber \\
    && m_{16}^{eq} = \rho u_xu_yu_z \ , \ \ m_{17}^{eq} = \rho c_s^2 \left [ 3(u_x^2u_y^2 + u_x^2u_z^2 + u_y^2u_z^2) + 2(u_x^2 + u_y^2 + u_z^2) + 1 \right] \ , \nonumber \\
    && m_{18}^{eq} = \rho c_s^4 \left [ 9(u_x^2u_y^2 + u_x^2u_z^2 - u_y^2u_z^2) + 6u_x^2 + 1 \right] \ , m_{19}^{eq} = \rho c_s^2 (u_y^2 - u_z^2)( 3u_x^2 + 1 ) \ , \nonumber \\
    && m_{20}^{eq} = \rho c_s^2u_yu_z(3u_x^2 + 1 ) \ , \ \ m_{21}^{eq} = \rho c_s^2u_xu_z(3u_y^2 + 1 ) \ , \ \ m_{22}^{eq} = \rho c_s^2u_xu_y(3u_z^2 + 1 ) \ , \nonumber \\
    && m_{23}^{eq} = \rho c_s^4u_x(3u_y^2 + 1)(3u_z^2 + 1 ) \ , \ \ m_{24}^{eq} = \rho c_s^4u_y(3u_x^2 + 1)(3u_z^2 + 1 ) \ , \nonumber \\
    && m_{25}^{eq} = \rho c_s^4u_z(3u_x^2 + 1)(3u_y^2 + 1 ) \ , \ \ m_{26}^{eq} = \rho c_s^6(3u_x^2 + 1)(3u_y^2 + 1)(3u_z^2 + 1 ) \ . 
\eea 

In the absence of a forcing term, the post-collision moments are given by $m_i^{*} = m_i^{eq}$ for $i = (0,...,3,9,...26)$ and
\bea 
    && m_4^{*} = (1 - \omega)m_4 + \omega\rho u_xu_y \ , \ \ m_5^{*} = (1 - \omega)m_5 + \omega\rho u_xu_z \ , \ \ m_6^{*} = (1 - \omega)m_6 + \omega\rho u_yu_z \ ,  \nonumber \\
    && m_7^{*} = (1 - \omega)m_7 + \omega\rho (u_x^2 - u_y^2) \ , \ \ m_8^{*} = (1 - \omega)m_8 + \omega\rho (u_x^2 - u_z^2) \ ,
\eea 

\noindent with
\bea
&& m_4 = \sum_i f_i c_{ix}c_{iy} \ , \ \  m_5 = \sum_i f_i c_{ix}c_{iz} \ , \ \  m_6 = \sum_i f_i c_{iy}c_{iz} \ , \ \nonumber \\
&& m_7 = \sum_i f_i (c_{ix}^2 - c_{iy}^2) \ , \ \  m_8 = \sum_i f_i (c_{ix}^2 - c_{iz}^2) \ . 
\eea

In the presence of an arbitrary external force, the last term in Eq.~(\ref{m*}) has to be considered, where the forcing term within the MRT collision approach, will be given by Eq.~(\ref{Rmrti}). Making use of the aforementioned equilibrium distribution (\ref{BoltzmannEquilibriumDistribution}), the forcing term will be given by

\bea 
    && R_0^{MRT} = 0 \ , \ \ R_1^{MRT} = \rho F_x \ , \ \ R_2^{MRT} = \rho F_y \ , \ \ R_3^{MRT} = \rho F_z \ , \ \ R_4^{MRT} = \rho (F_yu_x + F_xu_y) \ ,  \nonumber \\
    && R_5^{MRT} = \rho (F_zu_x + F_xu_z) \ , \ \ R_6^{MRT} = \rho (F_zu_y + F_yu_z) \ , \ \ R_7^{MRT} = 2\rho (F_xu_x - F_yu_y) \ ,  \nonumber \\
    && R_8^{MRT} = 2\rho (F_xu_x - F_zu_z) \ , \ \ R_9^{MRT} = 2\rho (\mathbf{F}\cdot \mathbf{u}) \ , \nonumber \\
    && R_{10}^{MRT} = c_s^2\rho \left[6 u_x (F_yu_y + F_zu_z) + F_x(2 + 3u_y^2 + 3u_z^2) \right ] \ , \nonumber \\
    && R_{11}^{MRT} = c_s^2\rho \left[6 u_y (F_xu_x + F_zu_z) + F_y(2 + 3u_x^2 + 3u_z^2) \right ] \ , \nonumber \\
    && R_{12}^{MRT} = c_s^2\rho \left[6 u_z (F_xu_x + F_yu_y) + F_z(2 + 3u_x^2 + 3u_y^2) \right ] \ , \nonumber \\
    && R_{13}^{MRT} = \rho \left[2 u_x (F_yu_y - F_zu_z) + F_x(u_y^2 - u_z^2) \right ] \ , \nonumber \\
    && R_{14}^{MRT} = \rho \left[2 u_y (F_xu_x - F_zu_z) + F_y(u_x^2 - u_z^2) \right ] \ , \nonumber \\
    && R_{15}^{MRT} = \rho \left[2 u_z (F_xu_x - F_yu_y) + F_z(u_x^2 - u_y^2) \right ] \ , \nonumber \\
    && R_{16}^{MRT} = \rho (F_xu_yu_z + F_yu_xu_z + F_zu_xu_y ) \ , \nonumber \\
    && R_{17}^{MRT} = 2c_s^2\rho \left [ F_xu_x( 2 + 3u_y^2 + 3u_z^2) + F_yu_y( 2 + 3u_x^2 + 3u_z^2) + F_zu_z( 2 + 3u_x^2 + 3u_y^2)  \right ] \ , \nonumber \\
    && R_{18}^{MRT} = 2c_s^2\rho \left [ F_xu_x( 2 + 3u_y^2 + 3u_z^2) + 3F_yu_y( u_x^2 - u_z^2) + 3F_zu_z(u_x^2 - u_y^2)  \right ] \ , \nonumber \\
    && R_{19}^{MRT} = 2c_s^2\rho \left [ 3F_xu_x( u_y^2 - u_z^2) + (F_yu_y - F_zu_z)(1 + 3u_x^2)  \right ] \ , \nonumber \\
    && R_{20}^{MRT} = c_s^2\rho \left [ (F_zu_y + F_yu_z)(1 + 3u_x^2) + 6F_xu_xu_yu_z  \right ] \ , \nonumber \\
    && R_{21}^{MRT} = c_s^2\rho \left [ (F_zu_x + F_xu_z)(1 + 3u_y^2) + 6F_yu_xu_yu_z  \right ] \ , \nonumber \\
    && R_{22}^{MRT} = c_s^2\rho \left [ (F_yu_x + F_xu_y)(1 + 3u_z^2) + 6F_zu_xu_yu_z  \right ] \ , \nonumber \\
    && R_{23}^{MRT} = c_s^4\rho \{ F_x (1 + 3u_y^2)(1 + 3u_z^2) + 6u_x\left [ F_zu_z(1+3u_y^2) + F_yu_y(1+3u_z^2) \right]  \} \ , \nonumber \\
    && R_{24}^{MRT} = c_s^4\rho \{ F_y (1 + 3u_x^2)(1 + 3u_z^2) + 6u_y\left [ F_zu_z(1+3u_x^2) + F_xu_x(1+3u_z^2) \right]  \} \ , \nonumber \\
    && R_{25}^{MRT} = c_s^4\rho \{ F_z (1 + 3u_x^2)(1 + 3u_y^2) + 6u_z\left [ F_yu_y(1+3u_x^2) + F_xu_x(1+3u_y^2) \right]  \} \ , \nonumber \\
    &&R_{26}^{MRT} = 2c_s^4\rho \{ F_xu_x (1 + 3u_y^2)(1 + 3u_z^2) + \nonumber \\ 
    && \hspace{1.7 cm} (1 + 3u_x^2)\left [ F_zu_z(1+3u_y^2) + F_yu_y(1+3u_z^2) \right]  \} \ .
\eea 

The post-collision populations are obtained from the post-collision moments by
\be
    \vert f^* \rangle = M^{-1}\vert m^* \rangle \ , \ \label{k*f*}
\ee

\noindent the collision step then ends and it is followed by the streaming step, as it was previously mentioned in Eq.~(\ref{streaming}).

\end{document}